\documentclass[fleqn,12pt,twoside]{article}
\usepackage{espcrc1}

\usepackage{graphicx}
\usepackage[figuresright]{rotating}

\newcommand{\be}{\begin{eqnarray}}
\newcommand{\ee}{\end{eqnarray}}

\def\ni{\noindent}

\newcommand{\AmS}{{\protect\the\textfont2
  A\kern-.1667em\lower.5ex\hbox{M}\kern-.125emS}}

\title{Observability of neutron stars with quarks}

\author{M. Prakash\address[SUNY]
	{Department of Physics and Astronomy, 
	State University of New York at Stony Brook, \\
	Stony Brook, New York 11794-3800, USA}, 
                J.M. Lattimer\addressmark,
                A.W. Steiner\addressmark\
		and
                D. Page\address{Instituto de Astronomia, UNAM, 
		Mexico D.F. 04510, Mexico}}
\begin{document}

% typeset front matter
\maketitle

\begin{abstract}

Recent {\em Chandra} observations have raised expectations that the
objects RX J185635-3754 and 3C58 are either bare strange quark stars
or stars with extended quark cores.  However, these observations can
also be interpreted in terms of normal neutron stars.  Essential
requirements for either explanation are that they simultaneously
account for the observed (i) spectral features (i.e., thermal or
nonthermal, possible lines), (ii) bounds on the inferred mass ($M$)
and radius ($R$), and (iii) the cooling curves (effective temperatures
$T$ and luminosities $L$ vs. age). Compact stars offer the promise of
delineating QCD at finite baryon density~\cite{LP01} in a fashion
complementary to relativistic heavy ion collider experiments, which
largely address the finite temperature, but baryon poor regime.
\end{abstract}

%\section{INTRODUCTION}
Recently, multi-wavelength photon observations of the isolated neutron
star RX J185635-3754 has attracted much attention since its distance
from Earth and hence its radius was estimated. The basis of a radius
estimation rests on relating the measured flux $f_\infty$ and
effective temperature $T_\infty$ (hereon, the subscript $\infty$
refers to quantites measured far away from the source) from a thermal
source of radius $R$, temperature $T$ and distance $D$ using
\be 
L_{\infty} = 4 \pi R_{\infty}^2 \sigma T_{\infty}^4 = 4 \pi D^2
f_{\infty} \qquad \Rightarrow \qquad R_{\infty} = D \sqrt{ f_{\infty} / (
\sigma T_{\infty}^4)} \,, 
\ee 
where $L_\infty$ is the luminosity.  The measured and in situ
quantities are related by appropriate red-shift factors: \be
R_{\infty} = \frac{R}{\sqrt{1-R_s/R}}, \qquad R_s=\frac{2GM}{c^2}
\qquad {\rm and} \qquad T_{\infty} = T {\sqrt{1 - R_s /R}} \,.  \ee In
practice, several additional considerations such as the gravitational
bending of light, nonuniform surface temperatures, redistribution of
flux through a neutron star atmosphere, and the effects of magnetic
fields, etc., must be included before the radius (and, in some cases,
the mass) of the neutron star can be estimated (see, e.g.,
Ref.~\cite{Pons02}). Table~\ref{table:1} contains a summary of
findings to date.  The column labelled ``THEN'' refers to the findings
of Ref.~\cite{Pons02} and is based upon the reported parallax in
Ref.~\cite{Walter01}, which, however, did not include corrections for
geometrical distortions of the HST camera at the edges of the fields.
Such corrections were incorporated in Refs.~\cite{KvKW02,WL02} with
the result that the parallax was reduced by roughly a factor of two.
Hence, the distance $D$ has been revised to nearly twice the value
quoted in Ref.~\cite{Walter01} (see the column labelled ``NOW'').

\begin{table}[hbt]
\caption{RX J185635-3754 
(ROSAT, EUVE, NTT, Keck, HST \& Chandra)}  
\label{table:1}
\renewcommand{\tabcolsep}{2pc} % enlarge column spacing
\begin{tabular}{lll}
\hline
 & THEN & NOW$^*$  \\
\hline 
Parallax (mas) & $16.5 \pm 2.3$ & $8.5 \pm 0.9$ \\  
Distance D (pc) & $60.6\pm 8.5$ & $117 \pm 12$ \\ 
Surface $T_{eff}^{\infty}$ (eV): & & \\ 
~~~~X-ray (Blackbody) & 57  & 63  \\
~~~~X-ray + Opt. + Atm. & $45\pm 6$   & $45\pm 6$  \\
Age (yr) & $0.9\times10^6$ & $5\times 10^5$ \\
Proper Motion (mas/yr) & $332 \pm 1$  & $332 \pm 1$  \\
Transverse Velocity (km/s) & $100\pm 15$ & $185\pm 26$ \\
$R_\infty=R/{\sqrt{1-2GM/c^2}}$ (km) & $6-8$ (BB) & $15\pm 3$ \\
$z=(1-2GM/Rc^2)^{-1/2}-1$ & $0.4\pm 0.1 $ & $0.35\pm 0.15$ \\
Mass $M/{\rm M}_\odot$ & $0.6-1.2$ & $1.7 \pm 0.4$ \\
Radius $R$ (km) & $4.5-8$ & $11.4 \pm 2$ \\ 
%Surface Field B (G) & ? & $\sim 10^{11}$ \\
Period P (s) & $-$ & $\sim 0.22$ \\  
\hline
\end{tabular}\\[2pt]
\ni $^*$ Includes 4th observation and corrections for geometrical
distortions of the HST camera at the edges of the fields.
\end{table}

\begin{figure}[hbt]
\begin{center}
\includegraphics[scale=0.5,angle=90]{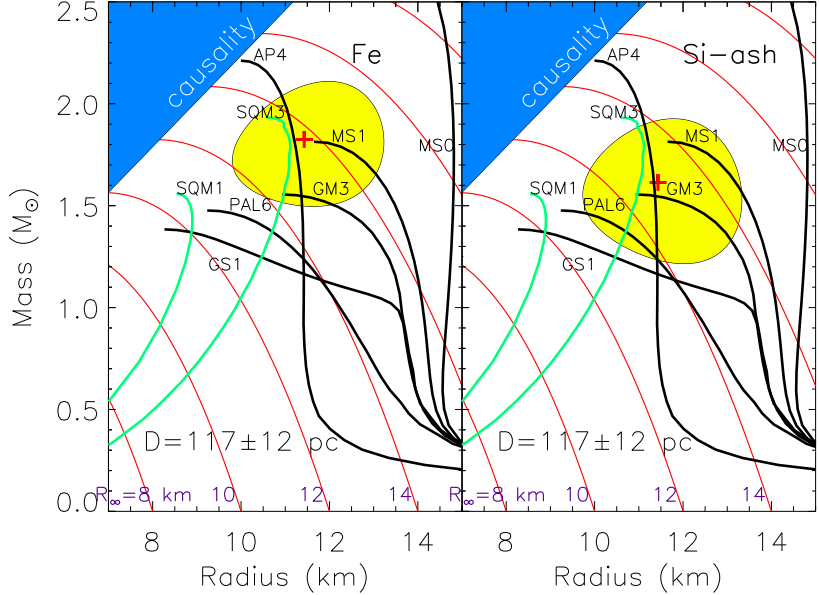} 
\end{center}
\vspace*{-0.65in}
\caption{Limits set by the RX J185635-3754 data in the mass ($M$)
versus radius ($R$) diagram. See text for details.}
\label{prakash_fig1} 
\end{figure}

Figure~\ref{prakash_fig1} shows mass-radius diagrams for
uniform-temperature heavy element atmosphere models, taken from
Ref.~\cite{Pons02}, for a revised distance of 117 pc. The left and
right panels are for Fe and Si-ash compositions, respectively.
Theoretical $M-R$ trajectories for representative equations of state
EOSs labelled following Ref.~\cite{LP01} are shown by thick solid
curves. The shaded region labelled ``causality'' is bounded by the
compactness limit set by requiring EOSs to be causal.  Thin lines are
contours of fixed $R_\infty$. The crosses denote the $M$ and $R$ 
of models which best fit the optical and X-ray data, and the
egg-shaped shaded regions surrounding them include the nominal errors
indicated in the constraint relations in equations (4) through (7) of
Ref.~\cite{Pons02}, as well as the nominal error in the distance.

The implications of the revised parallax and hence the distance are:
(i) The inferred radius ($R = 11.4 \pm 2$ km) and mass ($M/{\rm
M}_\odot = 1.7 \pm 0.4)$ are in the range of many EOSs both with and
without exotic matter, including cases of stars with quarks in their
core and bare strange quark matter stars, and (ii) Measurements have
removed observational support for an ``extremely soft'' EOS.  The
atmospheric composition and the effects of magnetic fields, however,
remain unresolved.

\begin{table}[hbt]
\caption{Pulsar J0205+6449 in 3C58 (Chandra)}  
\label{table:2}
\renewcommand{\tabcolsep}{2pc} % enlarge column spacing
\begin{tabular}{ll}
\hline
Period (ms) & $65$   \\
Energy loss $\dot E({\rm erg~s^{-1})} $ & $2.6\times 10^{37}I_{45}$ \\ 
Distance D (kpc) & $3.2$  \\
Column density $N_H$ (cm$^{-2}$) & $(3.75\pm0.11) \times 10^{21}$ \\
Power law index $\Gamma$ & $1.73\pm 0.07$ \\ 
Age (yr) & 922 (SN association) \\ 
         & 5481 (optical \& radio) \\
%$R_\infty=R/{\sqrt{1-2GM/c^2}}$ (km) & $12^*$ (BB) \\  
Surface $T_{eff}^{\infty}$ (eV) &  $\le 95^\dagger$ (BB) \\
\hline
\end{tabular}\\[2pt]
\ni $\dagger$ Data requires thermal component to be very small;
$T_{eff}^{\infty}$ is thus an upper limit. \\[2pt]
\end{table}

Recent Chandra observations of the pulsar J0205+6449 in the supernova
remnant of 3C58 indicate that the thermal component must be very
small, since the radiation is nearly completely fit by a power-law
spectrum~\cite{Slane02}. The temperature of a possible residual
thermal component is thereby limited (see Table~\ref{table:2}).
Although this upper limit can be fit with standard neutrino cooling
({\it e.g.,} $n+n \rightarrow n+p+e^{-}+\bar\nu_e$) plus pair-breaking
and formation, the luminosity and ages of other neutron stars cannot
be simultaneously fit using the same EOS~\cite{Yak02,Page02a}. This
has the interesting consequence that more exotic, rapid cooling
processes may exist in the neutron star core.

Among such processes are Urca reactions involving deconfined
quarks which may exist in many possible forms. These include a mixed
phase of hadrons and quarks, and color-flavor-locked (CFL) or 2-flavor
superconducting (2SC) or crystalline phases surrounded by an hadronic
phase with possible thin interfaces. It is also possibile that Bose
condensation may occur in the hadronic and/or CFL phases.  The most
exotic case would be a star made entirely of quark matter (the
so-called strange quark star) with CFL and/or 2SC phases with no
hadronic matter and therefore having so-called ``bare'' surfaces.

In Ref.~\cite{PPLS00}, the prospects of detecting baryon and quark
superfluidity from neutron stars during their long-term (up to $10^6$
years) cooling epoch was studied.  Our assessment is that future
photon observations of neutron star cooling (1) could constrain the
smaller of the $n-$ or $\Lambda-$ pairing gaps and the star's mass,
but (2) deducing the sizes of quark gaps will be difficult, (3) large
$q-$gaps would render quarks invisible, and (4) vanishing $q-$gaps
would lead to cooling behaviors indistinguishable from those of
ordinary nucleon or nucleon/hyperon stars.

In Ref.~\cite{JPS02}, neutrino emissivities from the decay and
scattering of Goldstone bosons in the color-flavor-locked (CFL) phase
of quarks at high baryon density were calculated.  Both emissivities
and specific heats are so small that, although the timescale for the
cooling of the CFL core bcomes exceedingly large, the CFL phase would
remain invisible because the exterior layers of normal matter
surrounding the quark core would continue to cool through
significantly more rapid processes.

In Ref.~\cite{Page02b}, it was noted that thermal emission from the
bare surface of a strange quark star can produce photon luminosities
well above the Eddington limit for extended periods of time, from
about a day to decades depending on the superconducting state of
strange quark matter. But the spectrum of emitted photons would be
significantly different from that of a normal cooling neutron star
($30 <E/{\rm keV} < 500$ instead of $0.1 < E/{\rm keV} <2.5$).  Due to
its distinctive spectrum and time evolution, such an observation would
constitute an almost unmistakable detection of a strange quark star
and shed light on color superconductivity at ``stellar''
densities. These predicted characteristics are well within the
capabilities of INTEGRAL~\cite{Integral} to be launched
towards the end of 2002.

The influence of quarks on neutrino fluxes from proto-neutron stars
were studied in Ref.~\cite{Pons01}.  Observable effects of quarks only
become apparent for stars older than 10--20 s.  Sufficiently massive
stars containing negatively-charged, strongly interacting, particles
(such as quarks, but also including hyperons and kaon condensates) 
may collapse to black holes during the first minute of evolution.
Since the neutrino flux vanishes when a black hole forms, this would
constitute an obvious signal that quarks (or other types of strange
matter) have appeared.  The collapse timescales for stars containing
quarks are predicted to be intermediate between those containing
hyperons and kaon condensates.


\begin{thebibliography}{99}
\bibitem{LP01}
J.M. Lattimer and M. Prakash, 
Astrophys. J. {\bf 550}, 426 (2001).  
\bibitem{Pons02}J.A. Pons, F.M. Walter, J.M. Lattimer, M. Prakash,
R. Neuhauser, and P. An, Astrophys. J. {\bf 564}, 981 (2002).
\bibitem{Walter01} 
F.M. Walter, Astrophys. J. {\bf 549}, 433 (2001). 
\bibitem{KvKW02} 
D. Kaplan, M. van Kerkwijk, and J. Anderson, 
Astrophys. J.  {\bf 571}, 447 (2002). 
\bibitem{WL02}
F.M. Walter and J.M. Lattimer, Astrophys. J. Lett., 
 {\bf 576}, L145-L148 (2002). 
\bibitem{Slane02}P. Slane et al.,  Astrophys. J.  {\bf 571}, L45, (2002).  
\bibitem{Yak02}D.G. Yakovlev et al., 
Astronomy \& Astrophysics, {\bf 389}, L24-L27, (2002).
\bibitem{Page02a}D. Page, J.M. Lattimer, M. Prakash, and A.W. Steiner, 
to be published. 
\bibitem{PPLS00}
D. Page, M. Prakash, J.M. Lattimer, and A.W. Steiner, 
Phys. Rev. Lett. {\bf 85}, 2048 (2000).
\bibitem{JPS02}P. Jaikumar, M. Prakash, and T. Sch\"affer, 
hep-ph/0203088, Phys. Rev. D., in print. 
\bibitem{Page02b}D. Page and V.V. Usov, Phys. Rev. Lett. {\bf 89},  
131101, (2002).  
\bibitem{Integral} See, e.g.,  astro-ph/0207527.
\bibitem{Pons01}J.A. Pons, A.W. Steiner, M. Prakash, and J.M. Lattimer, 
Phys. Rev. Lett. {\bf 80}, 230 (1998); 
Phys. Rev. Focus., June 1, 2001 (http://focus.aps.org/v7/st26.html). 
\end{thebibliography}
\end{document}